# Microfluidics for Ultra High-Throughput Experimentation: Droplets, Dots & Photons

*Andrew J. deMello - ETH Zürich*

**Biography**

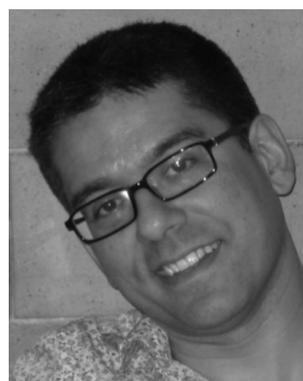

*Andrew is currently Professor of Biochemical Engineering in the Department of Chemistry and Applied Biosciences at ETH Zürich. Prior to this he was Professor of Chemical Nanosciences and Head of the Nanostructured Materials and Devices Section in the Chemistry Department at Imperial College London. He obtained a 1st Class Degree in Chemistry and PhD in Molecular Photophysics from Imperial College London in 1995 and subsequently held a Postdoctoral Fellowship in the Department of Chemistry at UC Berkeley working with Professor Richard Mathies. His current research interests cover a broad range of activities in the general area of microfluidics and nanoscale science, including the development of microfluidic devices for high-throughput biological and chemical analysis, ultra-sensitive optical detection techniques, microfluidic reaction systems for chemical and nanomaterial synthesis, the exploitation of semiconducting materials in diagnostic applications and the processing of living organisms. Andrew has given approximately 350 invited lectures at conferences and universities in North America, Europe, Africa and Asia (including 75 plenary/keynote lectures), has published over 300 papers in refereed journals, and co-authored two books. He currently sits on the Editorial Boards of Analytical Chemistry, The Journal of Flow Chemistry, Advanced Materials Technology and Chem. He is also co-founder of Molecular Vision Ltd, an Imperial College spin-out company developing low-cost diagnostic devices and Drop-Tech Ltd. Science originating from the deMello group has been recognized through the award of the 2002 SAC Silver Medal (RSC), the 2009 Clifford Paterson Medal (The Royal Society), the 2009 Corday Morgan Medal (RSC) and the 2007 Clark Memorial Lectureship (California State University). In 2012, Andrew was awarded the Pioneers of Miniaturization Lectureship by Dow Corning and the Royal Society of Chemistry.*

**Introduction**

The past 25 years have seen considerable progress in the development of microfabricated systems for use in the chemical and biological sciences. Interest in "microfluidic" technology has driven by concomitant advances in the areas of genomics, proteomics, nanoscale science, drug discovery, high-throughput screening and diagnostics, with a clearly defined need to perform rapid measurements on small sample volumes. At a fundamental level, microfluidic activities have been stimulated by the fact that physical processes can be more easily controlled when instrumental dimensions are reduced to the micron scale.[1] The relevance of such technology is significant and characterized by an array of features that accompany system miniaturization. These features include the ability to process small volumes of fluid, enhanced analytical performance, reduced instrumental footprints, low unit costs, facile integration of functional components and the capacity to exploit atypical fluid behaviour to control chemical and biological entities in both time and space.[2]

*A key theme of our work within the field has been the development of passive microfluidic systems that allow the performance of complex chemical and biological experiments on ultra-short timescales. To this end, we have spent significant effort in developing novel microfluidic systems for molecular and nanomaterial synthesis, droplet-based systems for ultra-high-throughput experimentation and novel optical techniques for sensitive and rapid analysis in small volume environments.*

## Droplet-Based Microfluidics

Droplet-based microfluidic systems (popularized by Ismagilov, Quake and Weitz) allow the generation and manipulation of discrete droplets contained within an immiscible carrier fluid.[3] Such systems leverage immiscibility to create discrete fL-nL volumes that reside and move within a continuous flow. Monodisperse droplets can be generated at rates in excess of tens of KHz, with independent control over each droplet in terms of its size, position and chemical payload. Significantly, the use of droplets in complex chemical and biological processing relies on the ability to perform a range of integrated, functional operations in high-throughput. Such operations include droplet generation, droplet merging/fusion, droplet sorting, droplet splitting, droplet dilution, droplet storage and droplet sampling.[4,5] We have used a range of passive droplet-based microfluidic systems (i.e. those in which droplet operations are effected through the variation of channel/feature geometries) to perform a variety of experiments inaccessible to both macroscale and continuous flow formats. Applications include intelligent nanomaterial synthesis,[6] cell-based assays,[7] high-throughput screening[8] and DNA amplification.[9] In all these areas, adoption of a droplet-based format transforms the efficiency of the reactive process or assay (through compartmentalization and confinement) and analytical throughput (due to the fact that tailored droplets may be produced at extremely high rates). It is significant to note that over the past decade, the development of droplet-based microfluidic technologies has occurred at a startling pace, with a focus on establishing functional operational components (for droplet processing) and pinpointing applications where the features of such systems may be used to the best effect. Based on their ability to perform complex experimental workflows in a rapid, efficient and robust fashion, the next decade will undoubtedly see the commercialization of many platforms for defined biological and chemical applications.

## Molecular and Nanomaterial Synthesis

Microfluidic systems have found significant utility in the broad field of molecular synthesis.[1] Put simply, the advantages associated with performing chemistry in microfluidic environments are significant and arise due to the scale-dependence of heat and mass transfer. These advantages, which include the ability to process reduced reagent volumes, improved reaction selectivity, the acceleration of mass-transfer limited reactions, small reactor footprints, enhanced safety and facile routes to scale-out, have made the technology set attractive as a synthetic tool. To this end, we have spent significant time in assessing the merits of microfluidic reactors.

For example, early studies from our group were the first to demonstrate small molecule combinatorial chemistry in flow-based environments.[10] Subsequently, we realized that continuous flow microfluidics should find significant utility in the controlled synthesis of nanoscale materials, due to the unique ability to temporally discriminate particle nucleation and growth processes. Our 2002 report of controlled quantum dot synthesis in a continuous flow microfluidic reactor demonstrated that CdS quantum dots prepared on the microscale exhibit superior monodispersity when compared to materials prepared in conventional macroscale reactors.[11] Since this initial publication we have detailed high efficiency routes for the synthesis of a range of nanomaterials, including CdS, CdSe, $TiO_2$, Au, $Fe_2O_3$, PbS, PbSe, $CsPbX_3$ perovskites, CdSeTe, $CuInS_2$, $CuInS_2$/ZnS, $Fe_3O_4$ and polymeric coacervates.[12] More recently, a key theme of our work in this area has been the development of "intelligent" systems for the generation of bespoke nanoscale materials. Briefly, microfluidic reactors integrated with real time product detection and a control system can perform and assay thousands of reactions within very short timeframes. We introduced such a strategy in 2007,[13] and since this time have used a variety of algorithms and metamodeling techniques to realize fully automated reactors for the production of a range of nanoscale materials.[6]

It should also be noted that we have used a range of chamber-based, continuous-flow and segmented-flow microfluidic formats for the amplification of nucleic acids via the polymerase chain reaction (PCR).[14,15,9] In all cases the enhanced mass and/or thermal transport that characterizes microfluidic processing leads to significant gains in terms of assay speed, analytical throughput and reaction efficiency.

## Small Volume Detection

The considerable advantages (with respect to speed, throughput, analytical performance, selectivity, automation and control) that are afforded through the use of microfluidic systems are in large part made possible by system downscaling and the associated improvements in mass and thermal transfer. Nonetheless, handling and processing fluids with instantaneous volumes on the femtoliter-to-nanoliter scale represents a critical challenge for molecular detection, and still defines one of the key limitations in the use of a microfluidic system in a given application. That is to say, although system downsizing leads to significant improvements in analytical performance, such gains are offset by a progressive reduction in the number of molecules which must be detected to report the chemical or biological state of the system. For example, it is well-recognized that the transfer of capillary electrophoresis to planar chip formats allows for ultra-short analysis times and high numbers of theoretical plates.[14,16] However, when performing electrophoretic separations within chip-based formats, analyte injection volumes are commonly no larger than 50 pL. This means that for an analyte concentration of 1 nM, only about 3000 molecules are available for separation and subsequent detection.

Over the past twenty years, we and others have addressed the issue of small volume analysis through the development of a range of detection methodologies. These include fluorescence and absorption spectroscopies, refractive index variation, Raman spectroscopies, infra-red spectroscopy, chemiluminescence, thermal lens microscopy and mass spectrometry. The pervasive adoption of optical methods is unsurprising, since chip-based microfluidic systems are typically fabricated from glasses, polymers and plastics, which have excellent optical transmission characteristics in the ultraviolet, visible and near infrared regions of the electromagnetic spectrum. Accordingly, molecular samples can be non-invasively probed via the absorption, emission or scattering of radiation.

Although the above methods are efficient at assaying small sample volumes on short timescales, they are typically inefficient at extracting the enormous amounts of chemical and biological information provided by microfluidic systems. Put simply, only a small fraction of the available information is commonly harvested during the detection process. To this end, a central theme of our activities in the area of small volume detection has been the development of novel optical methods that are able to perform multi-parameter analysis in a rapid an efficient manner. For example, we have pioneered the development of fluorescence lifetime imaging methods to probe temperature, pH, binding, viscosity and molecule concentration with unrivalled temporal and spatial precision.[17-19] We have also developed a range of vibrational and photothermal techniques for high sensitivity molecular fingerprinting.[20,21] These are of particular significance to chemical and biological systems, since they directly engender the high-throughput and sensitive analysis of non-fluorescent species, including small molecules, ions, biomolecule and nanomaterials. Most recently, we have developed entirely new ways of imaging rapidly moving species within microfluidic flows. These methods leverage stroboscopic illumination and allow the performance of enormous numbers of experiments on short timescales. Notable examples include the precision screening of over 1500 discrete enzymatic assays within 10 seconds,[22] and the demonstration of high-resolution multi-parametric imaging flow cytometry at unprecedented throughputs (in excess of 100,000 cells per second).[23]